\documentclass[10pt,journal,compsoc]{IEEEtran}
\usepackage{cite}
\usepackage{amsmath,amssymb,amsfonts}
\usepackage{algorithmic}
\usepackage{graphicx,color}
\usepackage{textcomp}
\usepackage{xcolor}
\usepackage{hyperref}
\usepackage{xurl}
\hypersetup{hidelinks}
\usepackage{algorithm,algorithmic}
\def\BibTeX{{\rm B\kern-.05em{\sc i\kern-.025em b}\kern-.08em
    T\kern-.1667em\lower.7ex\hbox{E}\kern-.125emX}}
\AtBeginDocument{\definecolor{tmlcncolor}{cmyk}{0.93,0.59,0.15,0.02}\definecolor{NavyBlue}{RGB}{0,86,125}}

\usepackage{xcolor}
\usepackage{colortbl}
\usepackage{array} 
\usepackage{tabularx}
\usepackage[inline]{enumitem}
\usepackage{graphicx}
\usepackage{booktabs}
\usepackage{tcolorbox}
\usepackage{tikz}
\usepackage{pifont}

\newcommand{\cmark}{\ding{51}} 
\newcommand{\xmark}{\ding{55}}   

\begin{document}




\title{Bad Diagnosis: Analyzing Consent and Data Collection Practices of Mobile Medical Apps}

\author{
\IEEEauthorblockN{
Ethan Myers\IEEEauthorrefmark{1},
Lorenzo De Carli\IEEEauthorrefmark{2},
Shantanu Sharma\IEEEauthorrefmark{3}
Ritwik Banerjee\IEEEauthorrefmark{4}
Joel Reardon\IEEEauthorrefmark{3}
Indrakshi Ray\IEEEauthorrefmark{1}
}
\IEEEauthorblockA{\IEEEauthorrefmark{1}Colorado State University, USA.}
\IEEEauthorblockA{\IEEEauthorrefmark{2}New Jersey Institute of Technology, USA. \\}
\IEEEauthorblockA{\IEEEauthorrefmark{3}University of Calgary, Canada.}
\IEEEauthorblockA{\IEEEauthorrefmark{4}Stony Brook University, USA.}
}

\maketitle






\begin{abstract}
Mobile health (mHealth) apps collect personally-identifiable information while handling uniquely sensitive medical data, ranging from mental health records to real-time glucose or heart rate monitoring---yet they remain poorly understood at scale.
We present the first comprehensive analysis of consent and data collection practices across 10,000+ mHealth apps on the Google Play Store, combining static analysis and privacy policy examination.
We develop a rigorous operational definition to distinguish health apps from general wellness 
 programs, and design a measurement framework to detect misuse of permissions and violations of consent.
Our analyses reveal a troubling ecosystem: 
\textbf{69.15\%} of apps request at least one permission unrelated to their stated purpose, with over \textbf{52.9\% apps requesting location data}, and \textbf{50.5\%} and \textbf{34.9\%} apps \textbf{requesting camera and microphone access}, respectively---despite no functional justification based on app descriptions and stated purposes.
Most alarmingly, nearly \textbf{93\% of apps fail to provide users with a consent form}, directly violating regulatory requirements such as GDPR; and even among the remaining 7\%, 1 in 14 demonstrate a gap between the obtained consent and collection of personally identifiable information.
These findings
demonstrate an urgent need for stronger enforcement of privacy regulations in mHealth apps. We plan to release our dataset and measurement tools for continued monitoring of this critical space.
\end{abstract}

\begin{IEEEkeywords}
Data privacy, information leakage, mobile applications, privacy 
\vspace{-0.25cm}
\end{IEEEkeywords}




\section{Introduction}

\label{sec:Introduction}

In April 2024, a mental health app was ordered to pay a \$7 million financial penalty for sharing sensitive data with third parties for advertising purposes without user consent between October 2019 and January 2023~\cite{ftc2024cerebal}. More recently, a federal class action lawsuit over a reproductive health app culminated in August 2025, when a jury found that background tracking disclosed sensitive health data to advertisers in violation of the app's own privacy policy~\cite{FrascovFlo}


These are not hypothetical risks, but documented failures in the mobile health (mHealth) ecosystem. Such failures point to systemic problems in the development and regulation of apps that promise to help users manage sensitive health information, while often mishandling it.
 Despite handling sensitive personal data and a market projected to be worth \$86 billion by 2030~\cite{mhealth_2025}, mHealth apps remain a uniquely dangerous point of failure for privacy violations.

These observations tie in with findings from past research on mobile apps in general, which lack rigorous security and privacy practices, including inconsistent communication to users and direct violations of privacy regulations~\cite{lapolla2013,yousra2024, alkindi2021, alloghani2020, delalamo2021}.
However, for mHealth apps specifically, these failures carry uniquely heightened risks: data leaks can lead to employment discrimination, insurance denial, or blackmail~\cite{wsj2024hackers}, presenting unique ethical, privacy, and trust challenges~\cite{ruotsalainen2020}.

In line with previous research~\cite{nguyen_freely_2022, nguyen_share_2021, koch_ok_nodate, kollnig_fait_2021, ardalani_precise_2024}, we focus on privacy violations at the point of data collection, i.e., app installation and initial usage.
This is when users decide whether to install the app, which data to provide, and which permissions to enable.
It is thus a critical juncture: once data has been provided, regaining control over it becomes infeasible~\cite{kaufman2011leakage, herrera2022survey}.
To rigorously identify and define privacy threats at this stage, we perform an analysis of apps' data lifecycle using the LINDDUN threat modeling framework~\cite{deng_privacy_2011}, following the domain filtering approach~\cite{Wuyts2018EffectiveAE}.
This analysis results in three privacy threats (PTs): the acquisition of permissions beyond an app's stated functionality (\textbf{PT1}), the absence of a mechanism to acquire user consent before data collection (\textbf{PT2}), and collecting data beyond acquired consent (\textbf{PT3}). Our PTs ensure comprehensive coverage of mHealth-specific privacy risks by avoiding narrow or arbitrary threat definitions noted in prior work~\cite{spensky2016,martin2016putting,ebrahimi2021,dev2023models}

Despite the severity of these risks, there is a dearth of prior work that comprehensively addresses all three privacy threats in mHealth apps.
Most existing studies either offer a broad examination of mobile apps (without mHealth-specific focus), or analyze only small samples of mHealth apps with limited scope.
More critically, large-scale analyses ($\geq$ 10,000 apps) have failed to measure the gap between acquired consent and actual data collection---leaving PT3 entirely unexamined at scale.
Similarly, prior measurements focusing specifically on mHealth apps have stopped short of collectively analyzing all three privacy threat types.

\begin{table}[!t]
\centering
\caption{Comparison with privacy measurement efforts on:
(i) focus on mHealth apps, 
(ii) analysis scale (${\geq}$10,000 apps or not), 
and (iii) examining PT1-PT3.} 
\vspace{-0.2cm}

\setlength{\tabcolsep}{3pt}
\scriptsize
\begin{tabular}{|c||c|c|c|c|c|}
\hline
\scriptsize
Study & Focus on mHealth & Large Scale & PT1 & PT2 & PT3 \\\hline
{\scriptsize  WHYPER (2013)~\cite{pandita_whyper_2013}} & \xmark & {\cellcolor[HTML]{bae4b3}{\cmark}} & \xmark & \xmark & \xmark \\
\hline
{AutoCog (2014)~\cite{qu_autocog_2014}} & \xmark & {\cellcolor[HTML]{bae4b3}{\cmark}} & {\cellcolor[HTML]{bae4b3}{\cmark}} & \xmark & \xmark \\
\hline
{ ACNET (2019)~\cite{feng2019acnet}} & \xmark & \xmark & \xmark & \xmark & \xmark \\
\hline
{ Feichtner et al. (2020)~\cite{feichtner_understanding_2020}} & \xmark & {\cellcolor[HTML]{bae4b3}{\cmark}} & {\cellcolor[HTML]{bae4b3}{\cmark}} & \xmark & \xmark \\
\hline
{  \textsc{HPDroid} (2020)~\cite{fan_empirical_2020}} & {\cellcolor[HTML]{bae4b3}{\cmark}} & \xmark & \xmark & \xmark & {\cellcolor[HTML]{bae4b3}{\cmark}} \\
\hline
{  Alecakir et al. (2021)~\cite{alecakir_attention_2021}} & \xmark & \xmark & \xmark & \xmark & \xmark \\
\hline
{  HybriDiagnostic (2021)~\cite{mohanty_hybridiagnostics_2021}} & \xmark & \xmark & \xmark & \xmark & \xmark \\
\hline
{  Tangari et al. (2021}~\cite{tangari2021mobile} & {\cellcolor[HTML]{bae4b3}{\cmark}} & {\cellcolor[HTML]{bae4b3}{\cmark}} & \xmark & \xmark & \xmark \\
\hline
{  \textsc{IoTSpotter} (2022)~\cite{jin_understanding_2022}} & \xmark & {\cellcolor[HTML]{bae4b3}{\cmark}} & \xmark & \xmark & \xmark \\
\hline
{  Neupane et al. (2022)~\cite{neupane_data_2022}} & \xmark & \xmark & \xmark & \xmark & \xmark \\
\hline
{  Nguyen et al. (2022)~\cite{nguyen_freely_2022}} & \xmark & {\cellcolor[HTML]{bae4b3}{\cmark}} & \xmark & {\cellcolor[HTML]{bae4b3}{\cmark}} & \xmark \\
\hline
{  Hassan et al. (2023)~\cite{hassan_what_2023}} & {\cellcolor[HTML]{bae4b3}\cmark} & \xmark & \xmark & \xmark & \xmark \\
\hline
{  Lyons et al. (2023)~\cite{lyons_log_2023}} & \xmark & {\cellcolor[HTML]{bae4b3}{\cmark}} & \xmark & \xmark & \xmark \\
\hline
{  PolicyChecker (2023)~\cite{xiang_policychecker_2023}} & \xmark & {\cellcolor[HTML]{bae4b3}{\cmark}} & \xmark & \xmark & \xmark \\
\hline
{  Ardalani et al. (2024)~\cite{ardalani_precise_2024}} & {\cellcolor[HTML]{bae4b3}{\cmark}} & \xmark & \xmark & \xmark & \xmark \\
\hline
{  Li et al. (2024)~\cite{li_are_2024}} & \xmark & \xmark & \xmark & \xmark & {\cellcolor[HTML]{bae4b3}{\cmark}}\\
\hline
{\bf {\cellcolor[HTML]{bae4b3}{This work}}} & {\cellcolor[HTML]{bae4b3}{\cmark}} & {\cellcolor[HTML]{bae4b3}{\cmark}} & {\cellcolor[HTML]{bae4b3}{\cmark}} & {\cellcolor[HTML]{bae4b3}{\cmark}} & {\cellcolor[HTML]{bae4b3}{\cmark}}\\
\hline
\end{tabular}
\label{tab:related}
\vspace{-0.35cm}
\end{table}

We close this knowledge gap by conducting the \textbf{first systematic analysis of mHealth apps at scale}, starting with the development of a rigorous operational definition of mHealth apps based on 
World Health Organization (WHO) guidelines \cite{who_mhealth_2018} and a highly accurate classifier---\textbf{94.6\% accurate with 0.92 F\textsubscript{1} score} (\S\hspace{1pt}\ref{sec:Dataset})---to identify \textbf{10,000+ mHealth apps} on the Google Play Store.
Based on the LINDDUN threat model, we distill and incorporate three types of privacy threats PT1--PT3 (\S\hspace{1pt}\ref{sec:Background}) into \textbf{an automated analysis pipeline}. This pipeline captures discrepancies between stated app functionality, permissions requested by the app, and its actual data collection behavior.
Our large-scale analysis methodology---described in \S\hspace{1pt}\ref{sec:methodology_pt2}, with results in \S\hspace{1pt}\ref{sec:results}\footnote{\scriptsize We make our dataset available at: \href{https://osf.io/j8zhd/?view_only=8fabbfd3ad0f41339e95e8a2c34631eb}{\texttt{https://tinyurl.com/bdcn9ta5}}}---reveals \textbf{pervasive privacy violations}: over \textbf{69\%} of the apps \textit{requested at least one permission unrelated to their stated purpose}, and nearly \textbf{93\%} \textit{lack a consent acquisition mechanism}, directly violating regulations like GDPR or CCPA.

\medskip 
\noindent \textbf{Why this paper?} This work is the first to perform a large-scale analysis of privacy implications of mHealth apps, using a comprehensive and rigorously defined threat model. Previous work (summarized in Table~\ref{tab:related} and further discussed in \S\hspace{1pt}\ref{sec:Background}) either lacks mHealth focus, only performs exploratory analysis on a handful of apps, misses important privacy risks, or a combination of these factors.

\vspace{-0.2cm}
\section{Background}
\label{sec:Background}
Like us, prior work largely concentrated on the Android ecosystem, owing to its open-source platform and widespread use, and may be broadly divided into two categories: measurement studies to quantify privacy risks, and threat modeling studies that develop frameworks to understand privacy threats.
The former usually focuses measuring common failure modes in mobile apps, such as the lack of consent, permission overreach, and deficient privacy policies~\cite{nguyen_freely_2022, nguyen_share_2021, pandita_whyper_2013, qu_autocog_2014, feng2019acnet, xiang_policychecker_2023, meng_large-scale_2024, ardalani_precise_2024}. On the other hand, the latter developed theoretical models and frameworks to aid our understanding of how privacy threats manifest in software systems~\cite{deng_privacy_2011, Wuyts2020LINDDUNGA}. 
These categories complement each other, and provide the foundation for the privacy threat types (\textbf{PT1}--\textbf{PT3}) analyzed in this work.

Table~\ref{tab:related} offers a comparison against the most closely related studies to highlight the gap bridged by this work: while some prior work achieved large scale or mHealth-specific focus, none have comprehensively measured all three privacy threat types at scale in the mHealth ecosystem. The work closer in spirit to us is Tangari et al.~\cite{tangari2021mobile}, which considered privacy properties of 15,000+ mHealth apps. Our work differs in two important way. First, Tangari et al. study is exploratory and has goals orthogonal to ours. Notably, this work is not concernend with measuring permission gaps (the focus of our PT1), or consent or violation thereof (PT2 and PT3). Second, this work has significant methodological issues, including: (i) scraping mHealth apps based on Play Store categories without further verification (in \S\hspace{1pt}\ref{sec:CollectionEffectiveness} we show doing so results in nearly 30\% FP); (ii) no support for analysis of TLS-encrypted traffic in the presence of certificate pinning (common in modern apps); (iii) inability to detect hashed or encoded PII; and (iv) use of external tools (policy labeling and PII detection) without prior validation on their dataset. Thus, we believe our work makes important contributions beyond what has already been investigated in the literature.

\vspace{-0.1cm}

\subsection{Scope: PII Categories}
We focus on 9 types of \textit{personally identifiable information} (PII) commonly collected by mobile apps (\autoref{tab:piitypes}), selected on the basis of established practices~\cite{lyons_log_2023, nguyen_share_2021}, and further refined for practical measurability. Our scope includes device identifiers like IMEI and Device Serial Number despite their restriction to system apps in Android 9+, allowing for investigations into potential side-channel collection attempts. On the other hand, we exclude PII categories that depend on carrier-specific configurations (e.g., SIM-based identifiers) or complex user interactions (e.g., account creation, automated logins), as these cannot be measured with static analysis or reproducible runtime monitoring, and would require advanced UI fuzzing beyond the scope of this work.

Having established which PII categories we examine, we now define the types of privacy threats that can arise at the point of mHealth app installation and permission acquisition---the critical juncture where users make trust-based decisions. 

\begin{table}[!t]
\centering
\caption{The 9 \textit{personally identifiable information} (PII) categories defining our scope, with brief descriptions.}
\vspace{-0.2cm}
\renewcommand{\arraystretch}{0.75}
\begin{tabularx}{\linewidth}{@{}l X@{}}
\toprule
  1. AAID & A unique, user-resettable, and user-deletable ID for advertising, provided by Google Play services~\cite{aaid_2025}\\
  
  2. Bluetooth MAC & The unique hardware identifier assigned to the device's Bluetooth.\\
  
  3. BSSID & The unique identifier for the WiFi access Point (AP) to which the device is connected.\\
  
  4. Device Serial Number & The device's serial number, provided by the manufacturer.\\
  
  5. Email & Email address associated with the device's Play Store account.\\
  
  6. IMEI & International Mobile Equipment Identity, the unique 15-digit identifier for the device.\\
  
  7. Nearby Bluetooth MAC & The unique hardware identifiers of nearby Bluetooth devices.\\
  
  8. SSID & The name of the WiFi network to which the device is connected.\\
  
  9. WiFi MAC & The unique hardware identifier for device's wireless network interface.\\
\bottomrule
\end{tabularx}
\label{tab:piitypes}
\vspace{-0.35cm}
\end{table}

Prior measurement studies of Android apps have documented common privacy failures including permission overreach~\cite{pandita_whyper_2013, qu_autocog_2014, feng2019acnet}, missing consent mechanisms~\cite{kollnig_fait_2021, nguyen_freely_2022}, and undisclosed data collection~\cite{meng_large-scale_2024}. These studies, however, typically examine individual failure modes in isolation rather than within a comprehensive privacy threat framework. In the next section, we discuss our approach to systematize the analysis based on the LINDDUN framework.

\vspace{-0.15cm}
\subsection{The LINDDUN Privacy Threat Framework}
To ground our analysis in a systematic framework, we adopt LINDDUN, a widely regarded privacy threat modeling framework introduced by Deng et al. in 2011~\cite{deng_privacy_2011}, and continually refined to address the evolving privacy landscape (e.g., the introduction of GDPR in 2018). LINDDUN, with its privacy-by-design framework, is geared to identify and mitigate privacy risks in software systems through the delineation of seven privacy threat categories:
\begin{itemize}[leftmargin=*, nolistsep, noitemsep]
\item \textit{Linking}: Associating data items or user actions to learn more about an individual or group.
\item \textit{Identifying}: Learning the identity of an individual, through leaks, deduction, or inference.
\item \textit{Non-Repudiation}: Being able to attribute a claim to an individual.
\item \textit{Detecting}: Deducing the involvement of an individual through observation.
\item \textit{Data Disclosure}: Excessively collecting, storing, processing or sharing personal data.
\item \textit{Unawareness}: Improperly informing, involving or empowering individuals in the processing of their personal data. 
\item \textit{Non-Compliance}: Deviating from security and data management best practices, standards, and legislation.
\end{itemize}
While LINDDUN offers comprehensive coverage of privacy threats, not all categories are relevant or measurable at scale for mHealth apps.
Following the \textit{domain filtering} approach~\cite{Wuyts2018EffectiveAE}, we focus on threats that emerge at the critical juncture of app installation and permission granting, which is also the point where users often decide on the basis of trust and limited information. 


We identify three privacy threats (PT1--PT3) that map to LINDDUN's categories. We describe these next with motivating examples, align them to prior work, and provide a succinct description of our approach. For further details of our methodology and improvements vis-\`a-vis each threat type, we direct the reader to \S\hspace{1pt}\ref{sec:methodology_pt2}.

\vspace{-0.2cm}

\subsection{PT1 (Permission Overreach)} 
PT1 is present when an app requests permissions that its stated functionality does not justify.

\noindent\textbf{Example:}
The app {\small\verb|com.transsion.healthlife|}, with over 163M downloads at the time of collection, requested permissions to access the device's camera, audio, storage, phone, call logs, and contacts---despite describing itself as a fitness tracking app that offers step counting, activity monitoring, and health tips. However, none of these features justify access to call logs or contacts, demonstrating clear permission overreach far beyond its purported functions.

\noindent\textbf{PT1 aligns with three specific LINDDUN threat types:}
Threats arising from limited comprehension of permission jargon and over-requesting of permissions align with these LINDDUN threat types:
\begin{itemize*}
\item[(1)] \textit{unawareness}, as the user may not understand the extent of permission usage and thus the data collection practices of the app;
\item[(2)] \textit{disclosure}, as over-requesting permissions leads to data collection beyond stated purposes; and
\item[(3)] \textit{non-compliance}, as over-requesting permissions leads to collection of data for which the user did not provide consent, thus constituting a lack of compliance.
\end{itemize*}

\noindent\textbf{Alignment to existing research:}
Android permission system serves as an (attempt to) implement the Principle of Least Privilege (PoLP) \cite{salterzer_computer_systems}, providing a mechanism to protect sensitive system resources.
But it cannot verify whether the requested permissions align with app functionality. This alignment remains at the discretion of the developer.
Prior work has investigated the misalignment between app descriptions and requested permissions using a variety of text mining and natural language processing (NLP) techniques, including keyword matching algorithms, semantic classifiers, and deep neural networks~\cite{pandita_whyper_2013, qu_autocog_2014, feng2019acnet}.

\noindent\textbf{Our approach:}
To measure if an app's requested permissions align with its claimed functionality, our approach is based on modern NLP: we fine-tune a large language model (LLM) as a multi-label classifier to predict which permissions are justified by an app's functionality, based solely on that app's Play Store description. 
We then compare these predictions against that app's actual requested permissions to identify overreach.
Our best model achieves a 99\% ROC-AUC on our test set (details in \S\hspace{1pt}\ref{sec:PT1methodology}).

\vspace{-0.15cm}
\subsection{PT2 (Absence of Consent Mechanism)} \label{sssec:sec2_pt2}
PT2 is present if the app does not have a consent mechanism.

\noindent\textbf{Example:}
The app {\small\verb|sixpack.sixpackabs.absworkout|}, with 177M+ downloads, collected the user's AAID (Google Advertising ID), without user consent, as the app lacked an action to present a consent dialog to the user.

\noindent\textbf{PT2 aligns with two specific LINDDUN threat types:}
From a threat modeling perspective, PT2 aligns with LINDDUN's threat types of
\begin{itemize*}
\item[(1)] \textit{unawareness}, and
\item[(2)] \textit{non-compliance}.
\end{itemize*}
The absence of a consent mechanism leaves users unaware of the data collection practices of these mHealth apps, leading to non-compliance within privacy regulations.

\noindent\textbf{Alignment to existing research:}
Previous large-scale measurement studies have documented numerous failures to obtain consent in Android apps~\cite{nguyen_freely_2022, nguyen_share_2021, kollnig_fait_2021, koch_ok_nodate}.
Large-scale analysis of generic Android apps showed that only a minority of them displayed a consent screen to users~\cite {nguyen_freely_2022}. Smaller exploratory works have found that third-party trackers commonly did not implement consent mechanisms~\cite{kollnig_fait_2021}, and consent dialogs often manipulated users into granting consent~\cite{koch_ok_nodate}. PT2's focus on consent is grounded by data protection law, which recognizes multiple lawful bases for data processing, including legitimate interest (e.g., GDPR Article 6(1)(f)~\cite{GDPR2016}). However, regulatory decisions consistently indicate that, in the context of behavioral advertising and tracking, consent is the only acceptable legal basis~\cite{EU_meta_v_others, art-29}. From this perspective, the quality and content of consent mechanisms are the decisive legal factor. Accordingly, PT2 concentrates on the presence of consent mechanisms rather than legitimate interest, aligning with previous research~\cite{nguyen_freely_2022, kollnig_fait_2021, koch_ok_nodate}.  


\noindent\textbf{Our approach:}
We launch each app in an automated testing environment. After an empirically determined waiting period to ensure proper app initialization and loading, we capture its initial UI screenshot.
We then extend a pre-existing privacy language dictionary---developed by Nguyen et al.~\cite{nguyen_freely_2022}---to identify consent-related language, resulting in 254 distinct consent forms.
Our manual inspection found only a single false positive---an accuracy of 99.6\% (details in \S\hspace{1pt}\ref{sec:PT2methodology}).

\vspace{-0.15cm}
\subsection{PT3 (Abuse of Consent)}
PT3 is present if some PII is being collected by an app without obtaining prior consent.

\noindent\textbf{Example:}
The \verb|com.oraimo.joywear2| app, with 1M+ downloads, collected users' AAID\@. However, their consent dialog did not include any statement about collecting it. Thus, the data collection was unauthorized and outside the scope of obtained consent.

\noindent\textbf{PT3 aligns with three specific LINDDUN threat types:}
Abuse of consent corresponds to three LINDDUNS threat types of
\begin{itemize*}
\item[(1)] \textit{unawareness}, as users are misled about actual collection practices; 
\item[(2)] \textit{non-compliance}, violating consent agreements and regulations like GDPR; and
\item[(3)] \textit{data disclosure}, through excessive collection beyond disclosed scope.
\end{itemize*}

\noindent\textbf{Alignment to existing research:}
Recent work has documented significant gaps between (a) data collection claims within consent dialogs and privacy policies, and (b) the actual data collection practices of mobile apps.
Notable findings include: the majority of apps have at least one data collection claim, but their privacy policies fail GDPR completeness~\cite{xiang_policychecker_2023}, third-party SDKs used by apps rarely provide privacy policies and often over-collect user data~\cite{meng_large-scale_2024}, and mobile health apps consistently leak data to third parties~\cite{ardalani_precise_2024}.

\noindent\textbf{Our approach:}
As PT3 pertains to the collection of PIIs beyond the scope of obtained consent, we examine only the subset of apps that provide a mechanism to obtain consent from their users.
We capture network traffic and logs during automated interactions to detect which PII types are collected by the app.
Separately, we train a classifier that, given a consent dialog, identifies PII collection claims.
Based on this classifier, we identify PT3 violations when the collected PII includes types not disclosed in the consent screen (implementation details in \S\hspace{1pt}\ref{sec:PT3methodology}).

\vspace{-0.15cm}
\subsection{Summary}
Prior work has described privacy failures: permission overreach~\cite{salterzer_computer_systems,
pandita_whyper_2013, qu_autocog_2014, feng2019acnet},
absence of consent~\cite{nguyen_freely_2022, nguyen_share_2021, kollnig_fait_2021, koch_ok_nodate}, and consent violations~\cite{xiang_policychecker_2023, meng_large-scale_2024, ardalani_precise_2024}.
However, these studies examine threats in isolation, and rarely have a specific focus on mHealth apps.
In \autoref{tab:related}, we summarized the key gaps in the most closely related research: existing large-scale studies lack mHealth-specific focus, while mHealth-specific studies examine only small samples or a subset of threats.
Our work addresses these gaps by defining and grounding PT1--PT3 in the LINDDUN framework, and systematically measuring them ``in the wild'', across 10,000+ mHealth apps.

\begin{figure*}[!t]
\vspace{-0.4cm}

  \centering
  \begin{minipage}{.275\textwidth}
    \centering
    \includegraphics[width=4cm,height=3.5cm]{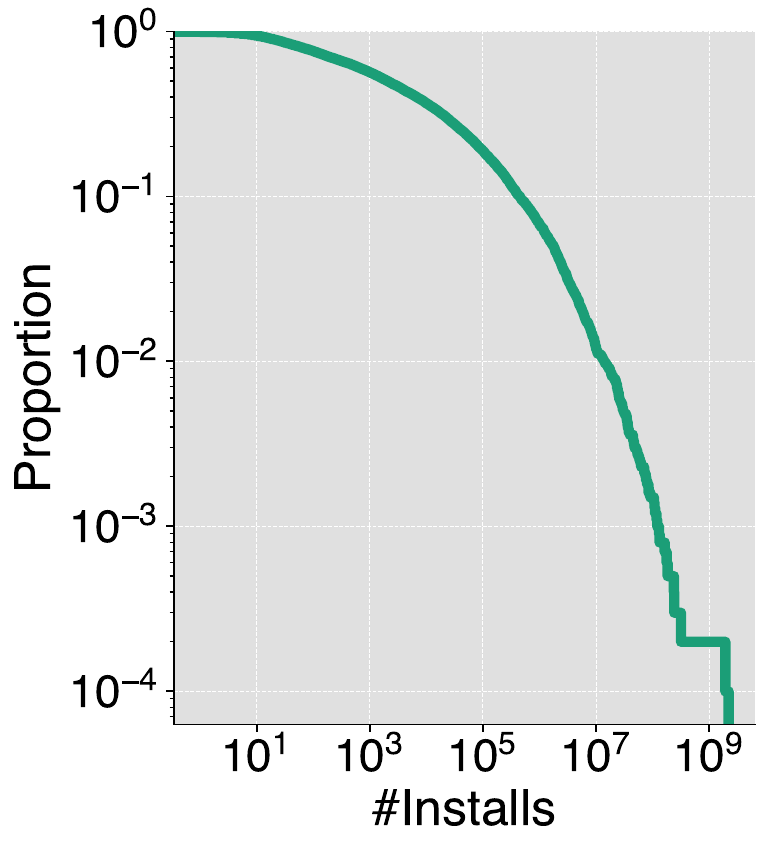}
  \end{minipage}
  \hspace{0.15in}
  \begin{minipage}{.275\textwidth}
    \centering
    \includegraphics[width=4cm,height=3.5cm]{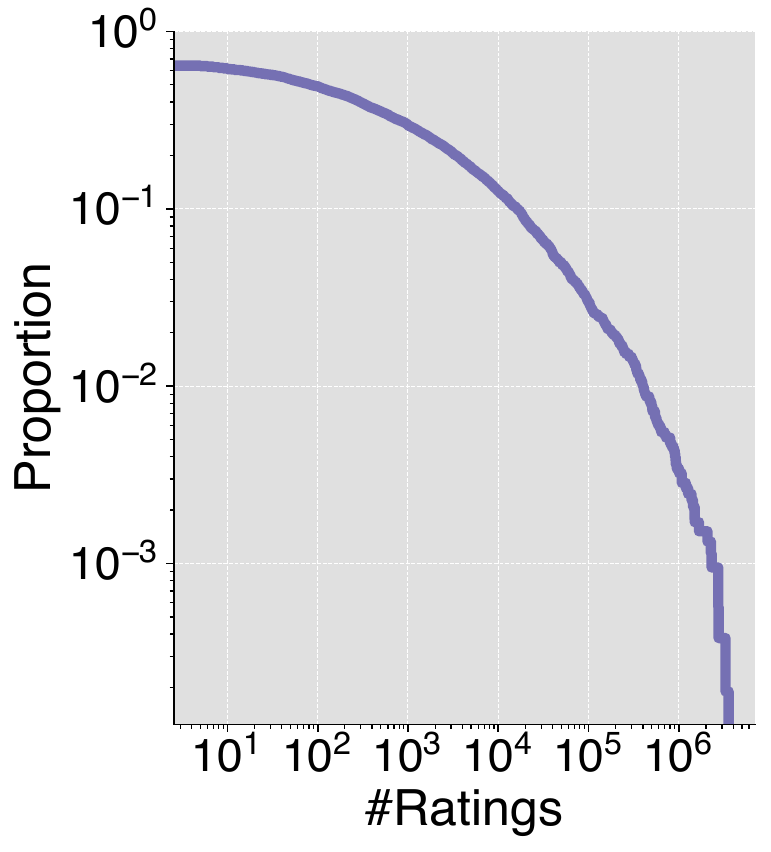}
  \end{minipage}
  \hspace{0.15in}
  \begin{minipage}{.275\textwidth}
    \centering
    \includegraphics[width=4cm,height=3.5cm]{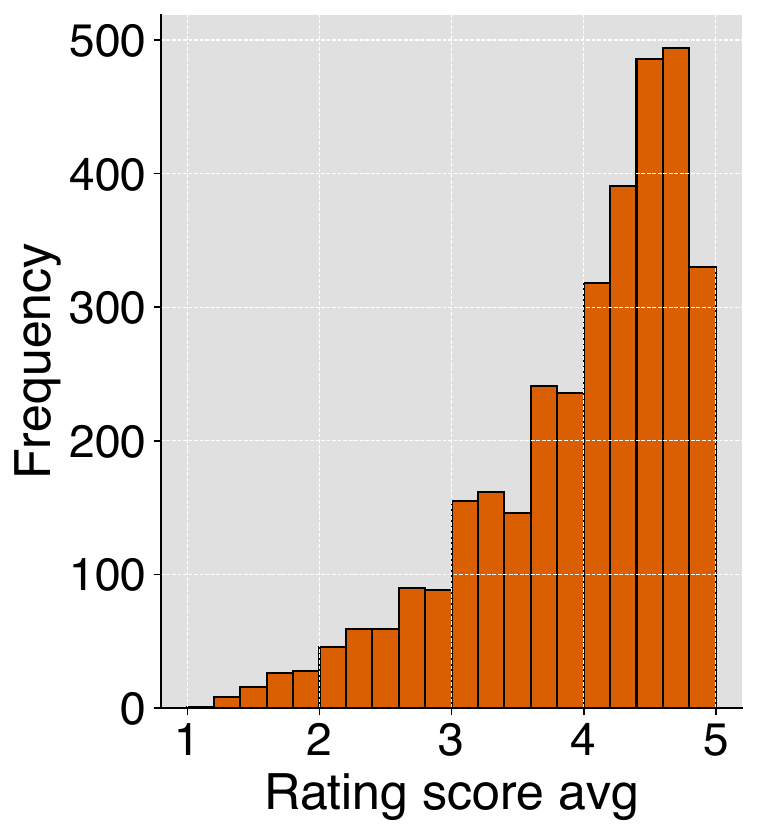}
  \end{minipage}
  \caption{Dataset characterization: The complementary cumulative distribution functions, or CCDFs, of app installs (left) and reviews (center) show heavy-tailed distributions; and the histogram of average rating scores (right) shows a significant positive skew toward higher ratings.}
  \label{fig:datasetchar}
  \vspace{-0.3cm}

\end{figure*}

\vspace{-0.15cm}

\section{Mobile Health (mHealth) App Dataset}

\label{sec:Dataset}
In this section, we discuss the process we followed to build our mHealth app dataset.
As we need to specifically analyze mHealth apps, automated data collection poses significant challenges:
\begin{itemize*}
\item[(i)] define \textit{what} constitutes an mHealth app (a question of scope), and
\item[(ii)] developing methods that offer coverage along with precision, for identification and collection of such apps at scale (a matter of design and implementation).
\end{itemize*}

\begin{table}[!t]
\centering
\setlength{\tabcolsep}{4pt}
\caption{The 5 best models for mHealth app curation (\S\hspace{1pt}\ref{ssec:mhealth-collection}), showing their (Acc)uracy and F$_1$, and the experiment settings: learning rate ($\eta$), number of epochs, and batch sizes.}
\vspace{-0.2cm}

\renewcommand{\arraystretch}{0.75}
\begin{tabularx}{\linewidth}{@{}Xccccc@{}}
\toprule
\textbf{Model} & \textbf{Acc.} & \textbf{F\textsubscript{1}} & $\mathbf{\eta}$ & \textbf{Epochs} & \textbf{Batch Size}\medskip\\
\rowcolor[HTML]{e5ffdd} DeBERTa (base) & 0.946 & 0.916 & 0.0002 & 3 & 16 \\
RoBERTa (base) & 0.941  & 0.909 & 0.0002 & 3 & 32 \\
DeBERTa (base) & 0.939 & 0.906  & 0.0002 & 3 & 32 \\
DeBERTa (base) & 0.939  & 0.905 & 0.0003 & 5 & 32 \\
DeBERTa (base) & 0.939 & 0.905  & 0.0002 & 5 & 32 \\
\bottomrule
\end{tabularx}
\label{table:top_5_models}
\vspace{-0.3cm}
\end{table}

\vspace{-0.15cm}
\subsection{An Operational Definition of \textit{mHealth Apps}}
\label{ssec:mhealth-app-definition}
Since there is no single universal definition of \textit{mHealth app}, we devise an operational definition based on criteria provided by the World Health Organization (WHO) for digital technologies for public health: viz., as software used in health-related services on mobile devices~\cite{who_mhealth_2018}.
These apps are meant to improve various aspects of health and healthcare and empower users to manage their health, with features such as health tracking, care support, secure data sharing with providers, privacy safeguards, and behavioral improvement strategies.
Based on this, we propose a taxonomy of mHealth apps derived from WHO guidelines, organized by their primary functional roles:
\begin{itemize}[leftmargin=*, nolistsep, noitemsep]
\item \textbf{Health Monitoring:} Apps designed to track and/or visualize physiological or behavioral data (e.g., heart rate, physical activity, blood pressure, sleep patterns).
\item \textbf{Disease Management:} Apps supporting chronic condition care (e.g., diabetes) through symptom logging, personalized dietary/exercise plans, or medication adherence tools.
\item \textbf{Telemedicine:} Apps facilitating remote clinical interactions (e.g., video consultations, secure messaging) for diagnosis, treatment, or follow-up care.
\item \textbf{Medication Management:} Apps tracking medication schedules, dosages, and interactions, often with adherence and safety alerts.
\end{itemize}
This framework distinguishes mHealth apps from general wellness tools by emphasizing their integration into healthcare workflows, data-driven decision-making, and adherence to clinical or regulatory standards. Despite overlaps (e.g., a telemedicine app may include health monitoring features), our definition prioritizes broad coverage and primary use cases to provide a pragmatic boundary.

\vspace{-0.15cm}
\subsection{Curation and Description of mHealth Apps}
\label{ssec:mhealth-collection}
Curating a large-scale dataset of mHealth apps that align with our definition presents significant challenges.
Existing approaches---including those relying on app store categories or keyword-based searches~\cite{ardalani_precise_2024, hassan_what_2023}---introduce substantial type I and type II errors: predefined categories often misclassify apps (e.g., delivery services identified as health apps because they deliver prescriptions or over-the-counter medicines), and keyword heuristics overlook relevant apps (e.g., the physiotherapy appointment app \textit{Mon Rdv Kin\'{e} - Praticien}, also called \textit{MyRedKine}, is labeled as ``productivity'').
These limitations underscore the need for a tailored collection framework.

Others have leveraged machine learning to glean specific types of apps from a large pool.
While some report low accuracy---and thus require manual intervention that cannot scale~\cite{wang_looking_2019, neupane_data_2022}---the use of Bidirectional LSTM and pretrained BERT embeddings~\cite{graves2005bilstm, devlin2019bert} for fine-tuning was demonstrably accurate in distinguishing IoT from non-IoT apps~\cite{jin_understanding_2022}.
We adopt a similar approach, but in contrast to their ensemble approach using consensus among two models, we find that fine-tuning a single model offers higher accuracy in identifying mHealth apps.

\begin{table}[!t]
\centering
\setlength{\tabcolsep}{8pt}
\caption{Summary statistics of mHealth user engagement and popularity indicators, including the number of times an app has been installed, average rating score, number of ratings, and number of reviews. Rows indicate the mean, standard deviation (Std), minimum (Min), 25th percentile (25\%), median (50\%), 75th percentile (75\%), and the maximum (Max) for each metric across our dataset.}
\vspace{-0.2cm}
\renewcommand{\arraystretch}{0.75}
\begin{tabularx}{\linewidth}{@{}Xrrrr@{}}
\toprule
& {\bf \#Installs} & {\bf Score} & {\bf \#Ratings} & {\bf \#Reviews}%
\medskip\\
\textbf{Mean} & 1.16M & 2.56 & 18,651.03 & 834.53\\
\textbf{Std} & 30.21M & 2.01 & 135,727.30 & 8,239.34\\
\textbf{Min} & 0 & 0 & 0 & 0\\
\textbf{25\%} & 111 & 0 & 0 & 0\\
\textbf{50\%} & 2,067.50 & 3.42 & 84.0 & 5\\
\textbf{75\%} & 42,914.25 & 4.39 & 1,790.50 & 61\\
\textbf{Max} & 2.20B & 5.0 & 3.56M & 449,419.0\\   
\bottomrule
\end{tabularx}
\label{tab:dataset_stats}
\vspace{-0.3cm}

\end{table}

\medskip
\noindent\textbf{The mHealth App Classifier.}
We constructed an annotated dataset of 5,000 apps to train our mHealth app classifier. To mitigate class imbalance, half the apps were sampled from the \textit{Medical} and \textit{Health \& Fitness} categories in Google Play Store, and the other half were randomly selected.
This sampling strategy balances the likelihood of including mHealth apps (via category filtering) with real-world diversity.
Two annotators independently labeled whether each app should be treated as an mHealth app on the basis of our operational definition.
After achieving near-perfect inter-annotator agreement (Krippendorff's $\alpha$ = 0.88-0.96) on two pilot batches of 100 apps each, the remaining 4,800 apps were annotated in parallel.
Upon completion of this process, we obtained 1,470 mHealth and 3,530 non-mHealth apps.

We fine-tuned transformer-based language models (BERT~\cite{devlin2019bert}, DeBERTa~\cite{he_deberta_2021}, RoBERTa~\cite{liu_roberta_2019}, and DistilBERT~\cite{sanh_distilbert_2020}) on this collection of 5,000 apps, optimizing hyperparameters for each architecture. DeBERTa achieved the best results (\autoref{table:top_5_models}: acc. = 0.946 and F$_1$ = 0.916), establishing it as our final classifier.

We used this classifier on 500,000 app descriptions from AndroZoo~\cite{allix_androzoo_2016}, identifying 10,688 mHealth apps.\footnote{\scriptsize We targeted 10,000+ apps to ensure sufficient sample size after validation and exclusion of paid apps, due to impracticality and cost of analyzing those.}
Between March--May 2024, we downloaded the corresponding APK files directly from the Play Store, restricting our collection to free apps to ensure broad accessibility and reproducibility.

\medskip\noindent\textbf{Comparison with category-based scraping.} 
\label{sec:CollectionEffectiveness}
We compare our classifier-based app collection pipeline with scraping based on Play Store categories, a simpler approach leveraged in past work~\cite{tangari2021mobile}. To do so, we select all apps belonging to Play Store Categories \textit{Medical} and \textit{Health \& Fitness} within our human-labeled dataset, and we test how many of them were labeled as mHealth apps by the annotators. Only 71.5\% of apps placed in the mentioned categories were labeled as mHealth apps, suggesting Play Store categories are insufficient for precise collection of mHealth apps.


\smallskip
\noindent\textbf{App Dataset Characterization}
\label{sssec:dataset-characterization}
\autoref{fig:datasetchar} characterizes our dataset through three measures: the left and center plots show complementary cumulative distributions (CCDF) of the number of app installs and the number of reviews on log-log scale, both exhibiting heavy-tailed distributions characteristic of large software ecosystems~\cite{vaidya_security_2021}, where a small fraction of apps account for the majority of user reach.
Together with the raw counts shown in \autoref{tab:dataset_stats}, this skew implies that the privacy practices of a few dominant apps disproportionately impact millions of mHealth users.
The third plot shows the distribution of average rating scores, which cluster around 4.5, suggesting positive bias in app store ratings.

\vspace{-0.2cm}
\subsection{Developer Geolocation}
\label{ssec:geolocation}
Geolocating app developers is critical for assessing regional regulatory impacts on privacy practices.
Developer locations are often incomplete or ambiguous, however (e.g., {\small\verb|com.wsl.noom|} lists a New York address but omits the country).
To address this, we designed a dedicated three-pronged country inference process:
\begin{enumerate}[leftmargin=*, nolistsep, noitemsep]
\item \emph{Direct Extraction:} Match app metadata to a dictionary of common country names and synonyms.
\item \emph{CrunchBase Query:} If the country is absent from the dictionary, query the CrunchBase company database API~\cite{crunchbase_2025}.
\item \emph{Fallback on top-level domains (TLDs):} Derive country names from the app package TLDs (e.g., {\small\verb|ca.teluspharmaconnect.mobile|} $\rightarrow$ Canada), excluding TLDs that are usually not geographic indicators (e.g., {\small\verb|.ai|} and {\small\verb|.io|}).
\end{enumerate}
A manual validation of 100 randomly geolocated apps confirmed 98\% accuracy; the two discrepancies stemmed from ambiguous public data. Overall, we geolocated 41\% of apps---a substantial coverage given that over 61\% did not specify any developer location.

While our sample may underrepresent regions with sparse developer location data, it sufficiently covers major markets (e.g., the EU, USA) with stringent privacy regulations. As such, this dataset enables analysis of regional privacy practices, even if partial.

\vspace{-0.2cm}
\section{Methodology} 
\label{sec:methodology_pt2}
We now discuss our methods to operationalize and measure the privacy threats (PT1--PT3) defined in \S\hspace{1pt}\ref{sec:Background}.

  \begin{table}[!t]
  \vspace{-0.1cm}
    \centering
    \caption{Comparison of BERT with prior work on permission inference from app descriptions~\cite{feng2019acnet,pandita_whyper_2013,qu_autocog_2014}.}
    \label{tab:model_performance}
    \vspace{-0.2cm}
    \renewcommand{\arraystretch}{0.75}
      \begin{tabularx}{\linewidth}{@{}>{\raggedright\arraybackslash}X r r@{}}  
      \toprule
      \textbf{Model / Paper} & \textbf{ROC-AUC} & \textbf{PR-AUC} \\
      \midrule
      BERT (this paper) & 0.99 & 0.81 \\
      AC-Net~\cite{feng2019acnet} & 0.97 & 0.66 \\
      Keyword-based~\cite{feng2019acnet} & 0.78 & 0.45 \\
      AutoCog~\cite{qu_autocog_2014} & 0.76 & 0.42 \\
      WHYPER~\cite{pandita_whyper_2013} & 0.60 & 0.32 \\
      \bottomrule
    \end{tabularx}
     \vspace{-0.3cm}
    \end{table}

\vspace{-0.2cm}
\subsection{PT1: Permission Overreach}
\label{sec:PT1methodology}
When apps request permissions that their stated functionality does not justify, it results in PT1. We operationalize this by 
comparing permissions implied by Play Store descriptions against permissions declared in app manifests. This difference constitutes the \textit{permission gap}.



\smallskip
\noindent\textbf{Permission Inference from App Descriptions.} We follow the approach taken by Tamrakar et. al~\cite{yunik_2025} by fine-tuning a language model to predict which permissions an app description justifies, framing this as multi-label text classification. 
Based on this approach, we build on the AC-Net dataset~\cite{feng2019acnet}, which contains 24,724 app description sentences, each labeled by human annotators with permissions that could reasonably be inferred from the text. 


\noindent\textit{Model Selection.}
We explored five pretrained models: BERT~\cite{devlin2019bert}, DistilBERT~\cite{sanh_distilbert_2020}, RoBERTa~\cite{liu_roberta_2019}, DistilRoBERTa~\cite{distilbertdistilroberta-base_2024}, and XLNet~\cite{yang2019xlnet}.
We focus on BERT-family models, as they achieve strong performance on text classification while being computationally more efficient than larger LLMs for our inference task~\cite{zhang_bert-like_2025}.
We fine-tuned each for 5 epochs using standard hyperparameters (learning rate $\eta = 2 \times 10^{-5}$, batch size 16), evaluating with two metrics: area under the receiver operating characteristic curve (ROC-AUC), and the precision-recall curve (PR-AUC), averaged over 10 runs.

\noindent\textbf{Addressing Class Imbalance.} The AC-Net dataset is highly imblanced; classical fine-tuning gave ROC-AUC (0.64)/PR-AUC (0.95), indicating the model struggled with minority classes. 
To address this, we generated synthetic paraphrases of minority-class samples using GPT-4o~\cite{gpt4o}, following the prompting strategy 
\cite{yunik_2025}.
After re-balancing the training set in this manner, we observed substantially improved performance (Table~\ref{tab:evaluation-metrics}).
Notably, the DistilBERT and DistilRoBERTa models matched the accuracy of full-sized models---despite having 35-40\% fewer parameters---thus enabling efficient inference.
We selected the fine-tuned BERT model for our analysis, based on its slight edge over DistilBERT in PR-AUC (Table~\ref{tab:evaluation-metrics}) and its superior performance over prior work in both metrics (Table~\ref{tab:model_performance}).

\noindent\textbf{Measuring PT1 ``in the wild''.} We sentence-tokenize app descriptions with NLTK and remove markup and emojis, lowercase, and run our BERT model with a 0.5 threshold. 
We extract the actual permissions by parsing app manifests using regular expressions. Because AC-Net only covers 16 permissions, PT1 is evaulated on that subset.  
The \textit{permission gap} for each app is the set difference: \texttt{(declared permissions) - (predicted permissions)}. PT1 is exhibited by mHealth apps with non-empty gaps.

\vspace{-0.2cm}
\subsection{PT2: Absence of Consent Mechanism} 
PT2 occurs when an app collects information but lacks any immediate mechanism to acquire user consent. To operationalize the measurement of this privacy threat, we base our approach on~\cite{nguyen_freely_2022}, which combines UI fuzzing and screenshot analysis to detect consent forms while monitoring for sensitive data transmission.

\label{sec:PT2methodology}

  \begin{table}[!t]
   \vspace{-0.5cm}
    \centering
    \caption{Evaluation metrics for top fine-tuned models.}
      \vspace{-0.1cm}
    \label{tab:evaluation-metrics}
    \renewcommand{\arraystretch}{0.75}
   \begin{tabularx}{\linewidth}{@{}>{\raggedright\arraybackslash}X r r@{}}
      \toprule
      \textbf{Model} & \textbf{ROC-AUC} & \textbf{PR-AUC} \\
      \midrule
      BERT (base-uncased)        & 0.990 & 0.815 \\
      DistilBERT (base-uncased)  & 0.991 & 0.808 \\
      RoBERTa (base)             & 0.990 & 0.815 \\
      DistilRoBERTa (base)       & 0.990 & 0.805 \\
      Class Distillation (XLNet) & 0.852 & 0.717 \\
      \bottomrule
    \end{tabularx}
     \vspace{-0.4cm}
  \end{table}
\smallskip
\noindent\textbf{Data Collection Environment}
Our data collection framework ran on three rooted Google Pixel 7a phones running Android 13. MITMproxy~\cite{mitmproxy} allowed us to capture network traffic, while we bypass SSL pinning using Objection~\cite{objection_2025}. We automated app interaction by first capturing a screenshot of the apps' initial screen, and then apply UI-fuzzing using Droidbot to explore different UI paths, trigger potenial data collection events, and grant all permissions. 




\noindent\textit{Critical modification} We identified several critical issues in the pipeline \cite{nguyen_freely_2022} and 
consequently engineered non-trivial adaptations to suit our work. Many apps failed to launch due to version incompatibilities or instrumentation, which we detected using OCR and string matching on captured screenshots. 
We also repaired corrupted screenshots by re-encoding the PNG meta data with SIPS~\cite{sips}. Lastly, we opted to use EasyOCR~\cite{easyocr} as our OCR engine as it empirally outperformed pytesseract~\cite{pytessaract}, which~\cite{nguyen_freely_2022} used.


\smallskip
\noindent\textbf{OCR model selection.} We benchmarked Python-tesseract ({\small\verb|pytesseract|})~\cite{pytessaract}
together with EasyOCR and TrOCR \cite{easyocr,li_trocr_2022} on a 500-sample subset of the 2019 SROIE dataset~\cite{huang_icdar_2019}. 
The Transformer-based TrOCR model outperformed all other models in terms of accuracy. However, it is limited to single-line text images \cite{li_trocr_2022}. We thus opted to use EasyOCR, as it can extract multiline text effectively.\footnote{\scriptsize Performing OCR on screenshots is a relatively easy task, and EasyOCR performance on our consent screenshots is significantly better than on the SROIE dataset.}

\smallskip
\noindent\textbf{Screenshot and Network Data Collection.} Our automated process installs each app and waits for 12 seconds, which we empirically determined to be enough time for the app to initialize and load its first screen. Then, it collects a screenshot of the screen displayed, and UI--fuzzes the app's interface for 150 seconds. Concurrently, we collect network packets from the app using MITMProxy. 
 Qualitatively, the UI--fuzzing algorithm used here leverages heuristics to attempt to elicit as many app behaviors as possible; however, it provides no guarantees about coverage. Note that, exhaustive UI path exploration is an open research problem~\cite{li_static_2017}. 
%
After filtering out crashed apps, we performed OCR to identify consent-related language within the captured screenshots, via matching against an extended version of the privacy language dictionary used by~\cite{nguyen_freely_2022}. 
We had to 
extend this to enable the accurate consent screen detection in regions beyond the EU, as \cite{nguyen_freely_2022} was GDPR-centric. We extended this dictionary by (i) identifying all 43 languages used in the descriptions of apps in our dataset (using Python {\tt \small langdetect} package); (ii) prompting the OpenAI GPT-4o model to extend the original dictionary to missing languages; and (iii) empirically verifying the correctness of a sample of wording for 10 languages using translation services.

\begin{table}[t]
 \vspace{-0.5cm}
\centering
\caption{OCR benchmark on SROIE \textnormal{(Huang et al.~\cite{huang_icdar_2019})}.}
   \vspace{-0.1cm}
\scriptsize
\renewcommand{\arraystretch}{0.75}
\begin{tabularx}{\linewidth}{>{\raggedright\arraybackslash}X r r r}
\toprule
\textbf{Model} & \textbf{Acc.(\%)} & \textbf{Time(s)} & \textbf{GPU Usage (MB)}\medskip\\
EasyOCR & 37.20 & 0.0674 & 160.34\\
Python-tesseract & 22.20 & 0.173 & (CPU-only) \\
TrOCR Base & 93.80 & 0.204 & 26.53 \\
TrOCR Small & 93.40 & 0.12 & 19.06 \\
\bottomrule
\end{tabularx}
\label{table:ocr-metrics}
     \vspace{-0.4cm}
\end{table}

Once screenshot analysis is complete, instances of PT2 can be identified by observing app screenshots without any consent-related language, while generated network data and/or logs contain PII (we discuss detection of PII in PT3).

     \vspace{-0.2cm}
\subsection{PT3: Abuse of Consent.} 
PT3 occurs when an app collects data beyond the consent it has obtained. We approximate PII collection by analyzing PII in each app's network communications and logs, and by creating (another) NLP pipeline over consent forms to determine whether the data collected was disclosed.   

\label{sec:PT3methodology}

\smallskip
\noindent\textbf{Collecting and Extracting PII from Network Data and System Logs.} Our collection and detection of PII is based on the approaches of~\cite{nguyen_share_2021} and~\cite{lyons_log_2023}. However, when building upon ~\cite{nguyen_share_2021} we found they failed to exclude non-app related traffic, which we rectified as described in the following. 

We search for a identifiers which include AAID, SSID, BSSID, email, IMEI, Bluetooth MAC, and device serial number in clear text along with encodings (e.g., base64), and transformations (e.g., MD5 and SHA-256) in both network traffic and logcat output. To avoid missattributing PII from apps that originated from Google Mobile Services (GMS) we build two heuristics. For logs, we identify logs that originated from the app through a combination of process identifier (PID) tracing and timestamp analysis. For network traffic, 
we use feature identification to identify characteristics of system generated traffic that include, request sent to known google domains (e.g., googleapis.com), very large payloads that reference internal android components (e.g., com.android.vending), protobuf or gRPC encoded binary, and no references to the package under testing. Across two manually labaled validation datasets, these hueristrics reduce false postives by 66.2\% in logs and 74.6\% in network traffic with 100\% accuracy.

\smallskip
\noindent\textbf{Categorizing and Extracting Consent Screens.}
When reviewing consent screens, we found three patterns: (i) 152 apps showed an embedded link to their privacy policy, (ii) 68 apps showed a short snippet of text, and (iii) 33 apps had text within a scrollable element. Due to this diversity, and the fact that case (iii) cannot be correctly extracted via screenshot+OCR only, we also attempted to extract consent forms from decompiled apps. To extract thier text, we use a two stage pipeline. First, we decompiled each app using JADX~\cite{skylot2025}, search for URLs and text resources associated with consent-related views, detect WebViews that load policy content, and search for specific phrases in the OCR'd text from the consent form screenshot (we found that using short phrases such as ‘privacy policy’ worked best). At the end of Stage 1, we had a list of URLs and associated files, dynamic calls, and text search results. In Stage 2, we manually searched the app's source code using the clues we had gathered in Stage 1. This process resulted in a 98.4\% successful consent screen extraction rate. We were unable to extract 4 consent screens due to decompilation errors. As a result, we fell back to OCR results. Overall, we have a total of 253 consent forms from our corpus.

\smallskip
\noindent\textbf{Data collection.} We construct a consent-form-text dataset by augmenting extracted consent form dialog with synthetic variants via GPT-4o, resulting in a corpus of 546 samples. Each sample is labeled based on whether the consent dialog mentions data collection for any of our PII identifiers. Two expert annotators labeled all samples, yielding a Krippendorff's $\alpha=0.765$. We then trained a sentence transformer 
to detect mentions of PII in consent form language.

We trained the sentence transformer paraphrase-multilingual-mpnet-base-v2~\cite{reimers-2020-multilingual-sentence-bert}, using SetFit~\cite{arxiv2209}, a few-shot contrastive learning method. We chose to use this model due to its high performance and because it was pre-trained on 50+ languages, making it suitable for non-English consent screens.  Our train and validation datasets were generated by selecting 20 types of PII from the majority labels and 30 from the minority PII types, e.g., IMEI, and Device Serial Number. 
Our validation dataset had 200 samples, and our training set had 219 samples. We selected 127 disjoint samples for the test dataset. We generate 30 contrastive pairs for each training sample using cosine similarity. Our model achieved 91.1\% F1 score and 97.5\% ROC AUC on our test set.

Finally, we measured the set difference between types of PII mentioned in consent form and PII collected within network data/logs, identifying instances of data collection beyond acquired consent.
\vspace{-0.2cm}
\section{Results}
\label{sec:results}
Our analysis of 10,688 mHealth apps reveals 
privacy violations across all three threat types: \textbf{69.15\% of apps exhibit permission overreach} (PT1), requesting permissions unjustified by their stated functionality.
Furthermore, nearly \textbf{93\% of apps lack any immediate consent mechanism} (PT2), indicating violations of regulatory requirements.
Among the apps that do obtain consent, \textbf{1 in 14 collect data beyond their disclosed scope} (PT3).
We also investigate whether popularity metrics are correlated with any threat type, as popularity and user opinion may influence (or be influenced by) good privacy practices.

\vspace{-0.2cm}
\subsection{PT1: Prevalence of Permission Gaps.}
PT1 measures the extent to which an app requests permissions that their stated functionality does not justify. We analyze 10,036 apps (94\% of our dataset), as the remaining 652 apps were removed from the Play Store between APK download and metadata collection.


\label{ssec:PT1measure}

\begin{table}[!t]
 \vspace{-0.2cm}
\centering
\setlength{\tabcolsep}{12pt}
\caption{PT1: Breakdown of permission utilization. No. of times permission is predicted (\#Pred.),  requested (\#Req.), and within the permission gap (\#In gap).}
\renewcommand{\arraystretch}{0.75}
\begin{tabularx}{\linewidth}{@{}>{\raggedright\arraybackslash}X r r r@{}}
\toprule
\textbf{Permission} & {\bf\#Pred.} & \ {\bf\#Req.} & {\bf\#In gap}\medskip\\
\textsc{camera}           & 2,206  & 6,477  & 5,062\\
\textsc{microphone}       & 2,216  & 4,434  & 3,502\\
\textsc{calendar}         & 1,701  & 3,324  & 2,809\\
\textsc{location}         & 8,670  & 10,914 & 5,303\\
\textsc{contacts}         & 9,152  & 3,718  & 2,205\\
\textsc{storage}          & 15,412 & 14,629 & 3,068\\
\textsc{phone}            & 429    & 838    & 722\\
\textsc{settings}         & 2,145  & 279    & 184\\
\textsc{tasks}            & 2,582  & 286    & 118\\
\textsc{call\_log}        & 88     & 209    & 154\\
\textsc{sms}              & 1,144  & 206    & 35\\
\bottomrule
\end{tabularx}
\label{tab:permissions}
   \vspace{-0.3cm}
\end{table}

\begin{figure*}[t]
     \vspace{-0.45cm}
     \centering
  \begin{minipage}[t]{0.32\textwidth}
    \centering
    \includegraphics[scale=0.24]{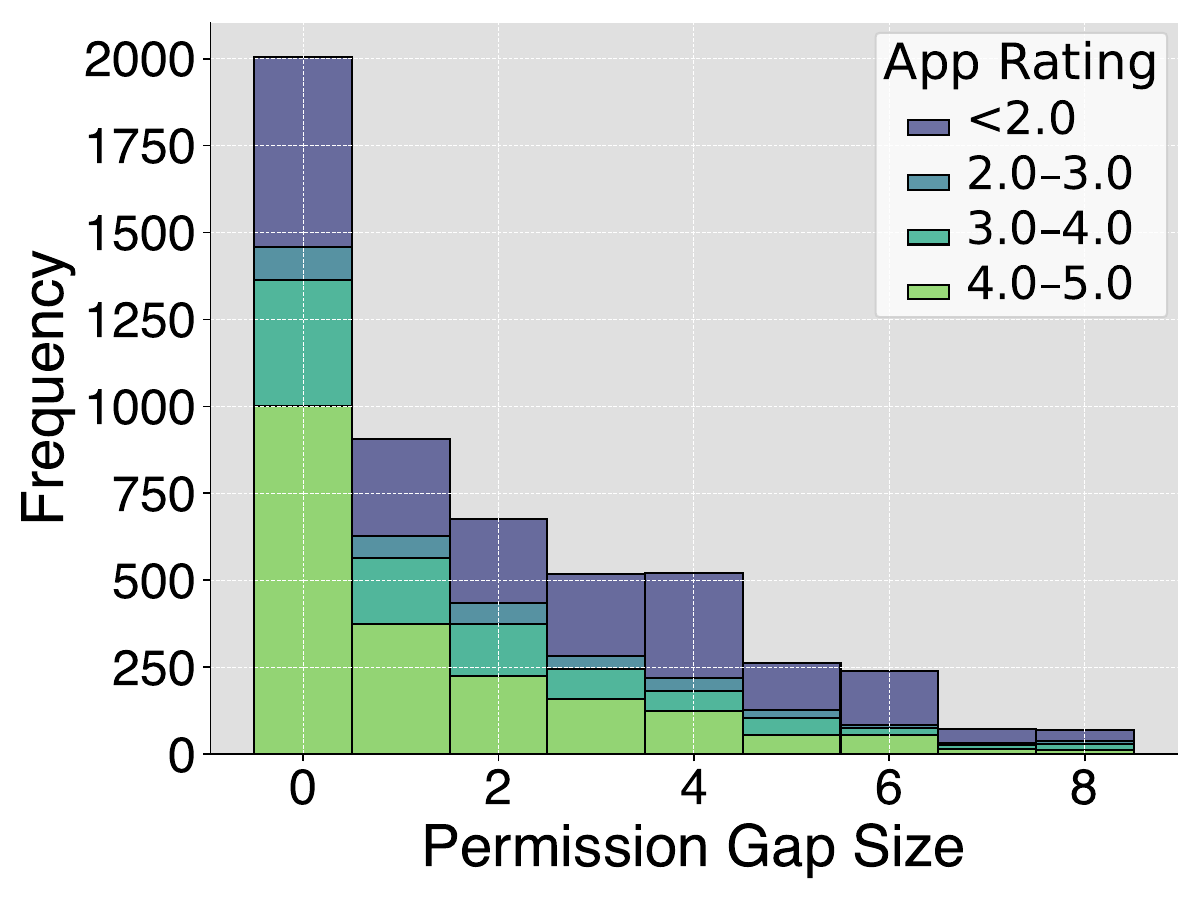}
  \end{minipage}
  \hfill
  \begin{minipage}[t]{0.32\textwidth}
    \centering
    \includegraphics[scale=0.24]{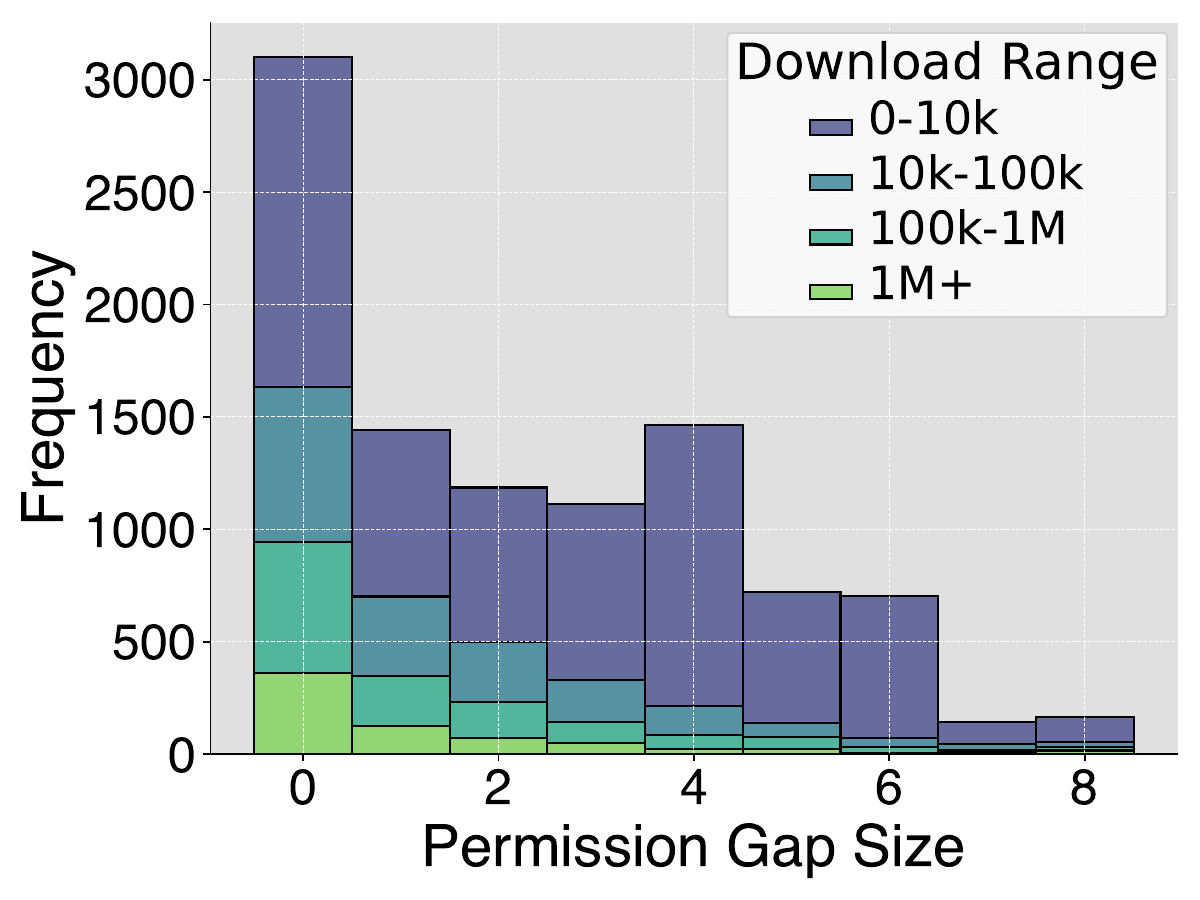}
  \end{minipage}
  \hfill
  \begin{minipage}[t]{0.32\textwidth}
    \centering
    \includegraphics[scale=0.24]{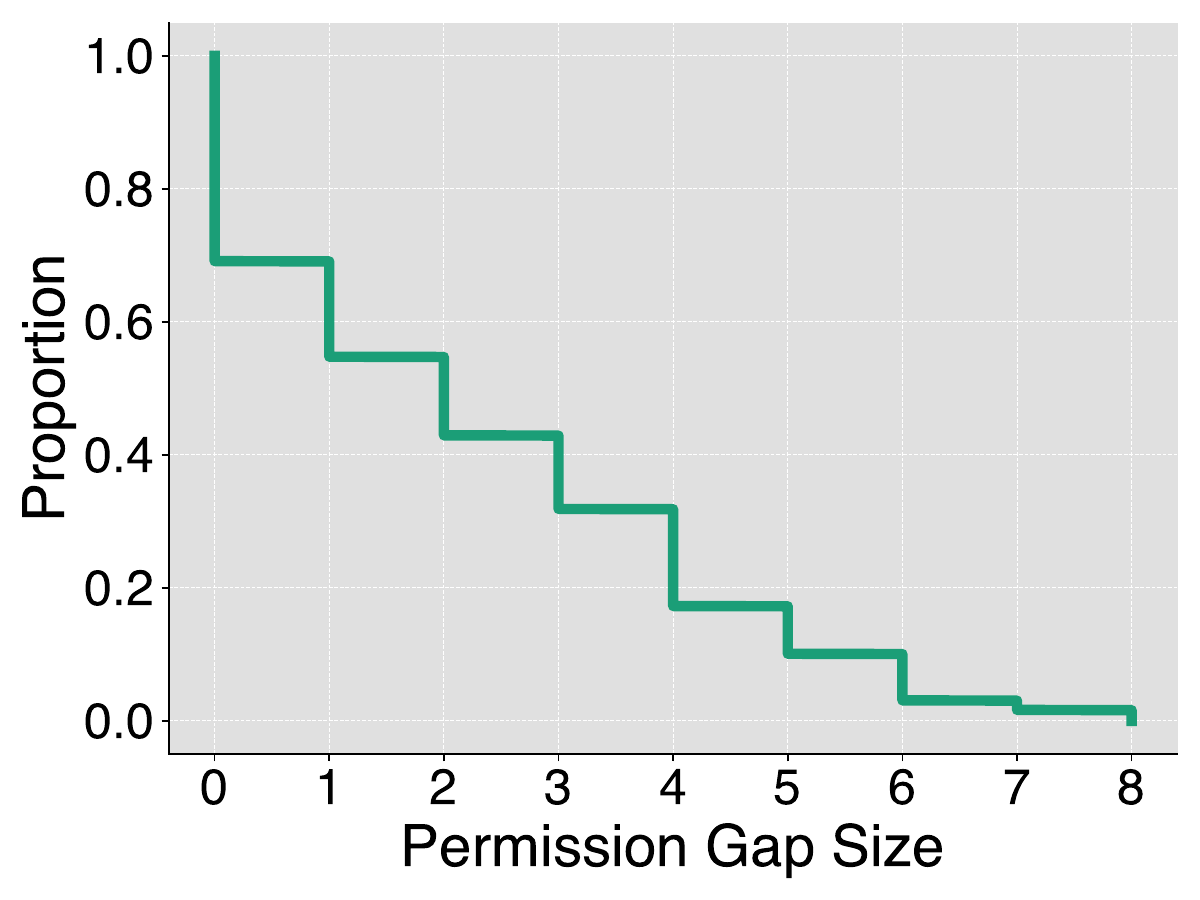}
  \end{minipage}
  \caption{PT1: Permission gap characteristics: the average permission gap by ratings (left) and installs (center), and the empirical cumulative distribution function, or eCDF, of permission gap sizes (right).}
  \label{fig:pt1-permission-gap}
     \vspace{-0.35cm}
\end{figure*}

Table~\ref{tab:permissions} shows the counts for the 11 permission categories: how often each permission is (i) requested in app manifests, (ii) predicted as necessary based on descriptions, and (iii) found in the \textit{permission gap} (i.e., requested but not predicted).
Among the 10,036 analyzed apps, 69\% (6,936 apps) exhibit permission gaps, requesting at least one permission their descriptions do not justify. The median gap size is 2 permissions (inter-quartile range, IQR = 4),
with most excess permissions providing access to sensitive user data such as location, camera, and microphone. The majority of permission gaps had multiple permissions, as depicted in Figure ~\ref{fig:pt1-permission-gap} (right). This sets a concerning precedent that could allow these apps to collect far more data than what the user is informed of from multiple data sources.

Predicted permissions at times exceeded requested permissions.
This may happen if app descriptions imply they have functionality they do not have, e.g., \texttt{com.terraillon.wellnesscoach} description suggest the use of \texttt{SMS} (``...receive call  sms notifications on your device ... '') without requesting such permissions. Also, we leveraged the labeling schema proposed by \cite{feng2019acnet}, where one permission label maps to multiple Android permissions (e.g., \texttt{CONTACTS} maps to \texttt{GET\_ACCOUNTS,~READ\_CONTACTS}).


\begin{figure}[!b]
   \vspace{-0.3cm}
\centering
\includegraphics[scale=0.08]{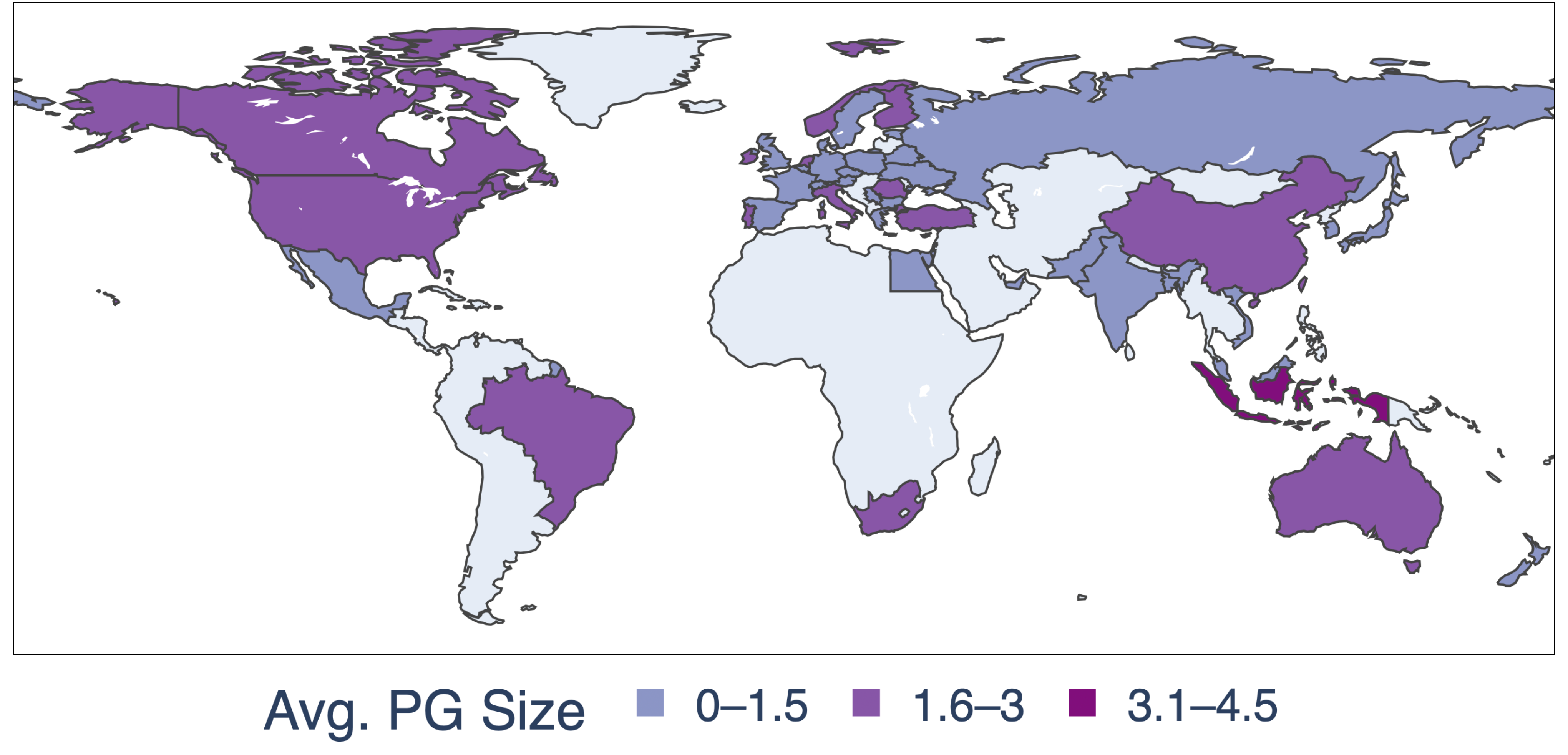}
\caption{PT1: Average permission gap size by country, excluding countries with less than 10 apps (in gray).}
\label{fig:pt1-geoplot}
\end{figure}

\noindent\textbf{Geographic Patterns}
Using developer location data, we geolocated 41\% of apps to analyze geographic patterns.
\autoref{fig:pt1-geoplot} shows the average permission gap sizes by region.
Surprisingly, apps from regions with existing privacy regulations---including the EU, South America, and parts of Asia---exhibit large average gaps (3.1-4.5 excess permissions), suggesting that regulatory frameworks alone may not ensure adequate user disclosure. Among these regions, Central and Southeast Asia show the largest average gaps. These findings are, however, limited by incomplete geolocation coverage and potential confounding factors such as developers being based elsewhere than their target markets.

\noindent\textbf{Popularity and User Experience Analysis.}
To investigate how app popularity (measured in terms of the number of app installs) varies with the permission gap, we binned the apps by their number of installs, using the bins illustrated in \autoref{fig:pt1-permission-gap} (center).
We find that the most highly downloaded apps have smaller permission gaps: apps with 1M+ installs have an average permission gap of 1.24, while less popular apps---with 100K-1M, 10K-100K, and <10K installs---have higher permission gaps of 1.39, 1.63, and 2.79, respectively.
To test whether these differences are significant, we conduct a Kruskal-Wallis test (since a Shapiro-Wilk test revealed the distribution to be non-Gaussian) to find statistically significant differences across the 4 bins ($H(3, N=10009)=958.60; p<0.0001$).
We then further conduct post-hoc pairwise Dunn's test with Bonferroni correction, and find that all pairwise differences are statistically significant except the 1M+ and 100K-1M installs.
%
%
A similar, but weaker, trend exists between permission gap size and average rating score (\autoref{fig:pt1-permission-gap}, left).
This is indicated by the Spearman correlation ($\rho = -0.025, p < 0.01, 95\% \text{CI} [-0.027, -0.022]$).
Given the small effect size and numerous confounding factors (app quality, functionality, user expectations), we leave further investigation of this latter result to future work.

\vspace{-0.2cm}
\subsection{PT2: Absence of Consent Mechanism}

We detect immediate consent mechanisms by launching each app, capturing its initial screen, and identifying consent-related language (see \S\hspace{1pt}\ref{sec:PT2methodology} for details).
Due to app crashes and execution failures, we successfully analyze 3,599 apps (34\% of our dataset)---a limitation we further discuss in \S\hspace{1pt}\ref{sec:discussion-and-limitations}.
Of the 3,599 analyzed apps, only 253 (7\%) displayed any consent mechanism to the user. \textbf{The remaining 93\% mHealth apps lack immediate consent acquisition.} {Among these apps, they collected
356 AAID, 1 BSSID, 25 Email, 1 SSID, 5 Bluetooth MAC, and 2 WiFi
MAC.} These findings align with prior work showing low consent rates (5.5\%) and non-compliant data collection across 
general Android apps~\cite{nguyen_freely_2022} and 
indicates that these failures extend to the privacy-critical mHealth domain.

\label{ssec:PT2measure}




\begin{figure}[!b]
 \vspace{-0.3cm}
\centering
\includegraphics[scale=0.22]{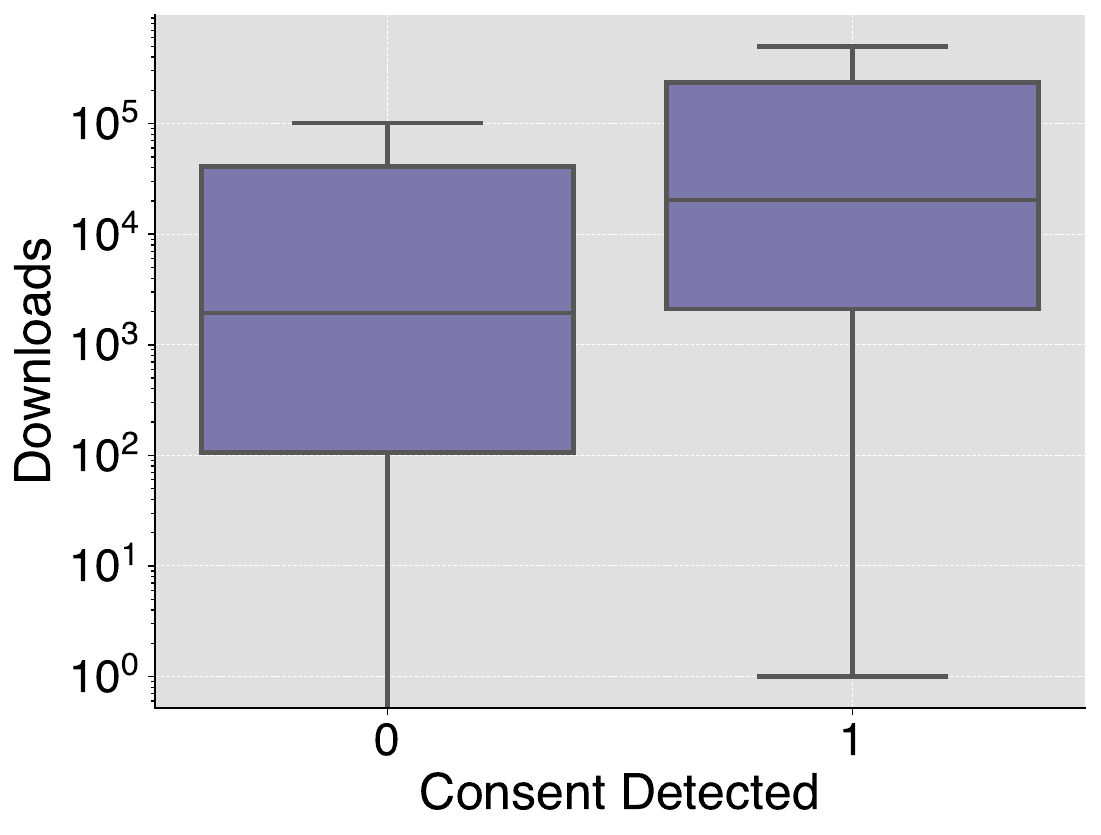}
\vspace{-0.15cm}
\caption{PT2: Distribution of downloads across presence/absence of immediate mechanisms to obtain user consent (excluding outliers).}
\vspace{-0.1cm}
\label{fig:pt2-boxplots}
\end{figure}

\begin{figure}[!b]
\centering
\includegraphics[scale=0.08]{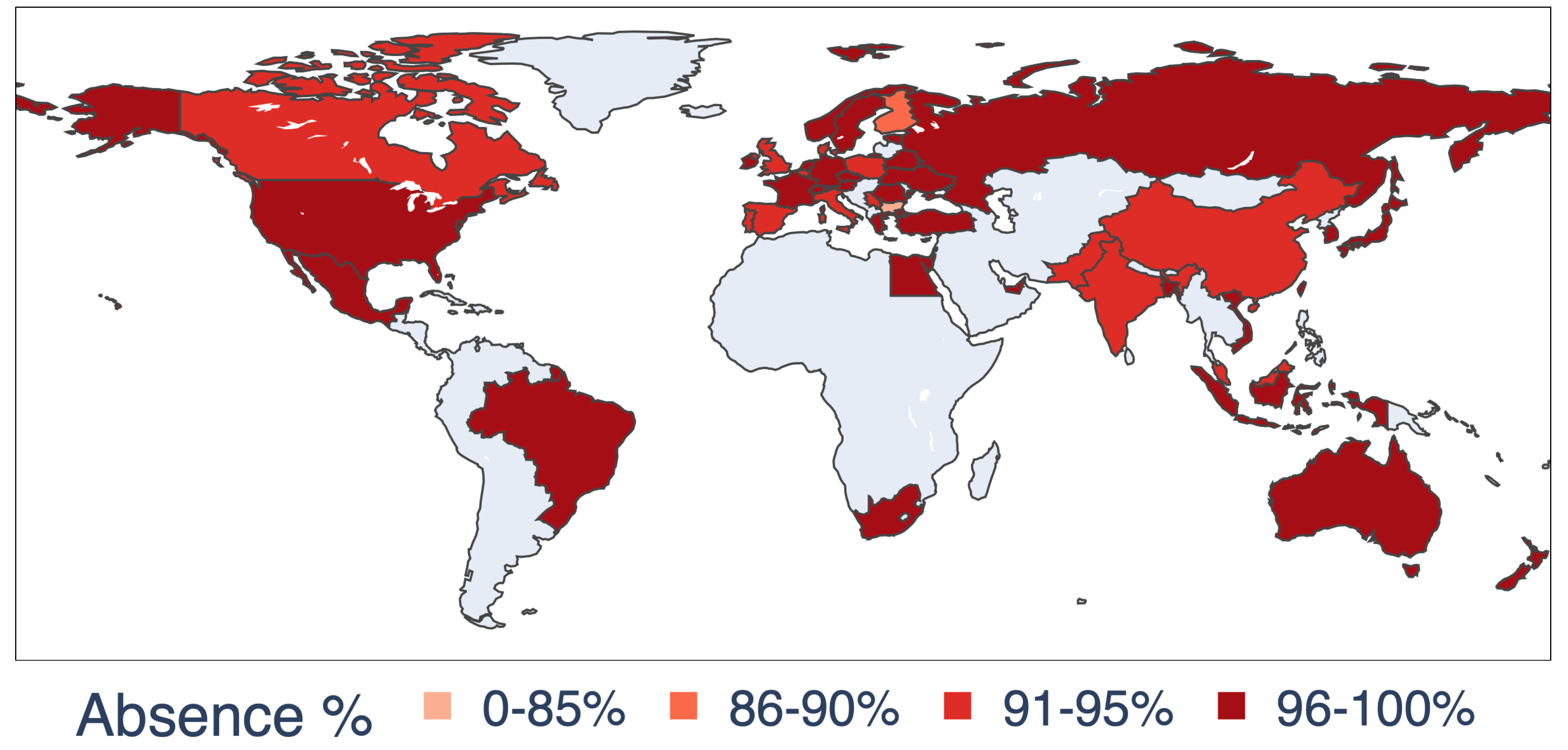}
\vspace{-0.3cm}
\caption{PT2: Absence of consent percentage by country, excluding countries with less than 10 apps (in gray)}
  \label{fig:pt2_geoplot_new}
\end{figure}

\noindent\textbf{Geographic Patterns.}
Among the 253 apps with consent mechanism, we geolocate 121 (48\%) to analyze regional patterns.
\autoref{fig:pt2_geoplot_new} and \autoref{tab:consent_by_region} show the distribution of apps that fail to offer consent mechanisms, across countries and (excluding countries with $<$10 apps) continents, respectively.
While this sample is limited, it indicates consistently low consent rates across regions, including those with established privacy regulations. However, strong conclusions about geographic patterns are not warranted, given the small sample size and high failure rate.

\begin{table}[t]
 \vspace{-0.2cm}
\centering
\caption{PT2: Percentage of apps, by region, with consent mechanisms. The rate (\%) indicates the proportion of apps in each region that provide users with \textit{immediate} (i.e., in the first screen of the app) mechanisms to manage their data consent preferences.}
\label{tab:consent_by_region}
\renewcommand{\arraystretch}{0.75}
\scriptsize
\begin{tabularx}{\linewidth}{@{}>{\raggedright\arraybackslash}X r r r r@{}}
\toprule
\textbf{Region} & \textbf{Apps w/ Consent} & \textbf{Total Apps} & \textbf{Rate (\%)}
\medskip\\

Africa &  1  &  57 & 1.75\\
Asia   & 27  & 797 & 3.38\\
South America &  3 &  142 & 2.11\\
Europe        & 53 & 1,707 & 3.10\\
Oceania       &  4 &  141 & 2.83\\
North America & 33 & 1,264 & 2.61\\
\bottomrule
\end{tabularx}
\vspace{-0.5cm}
\end{table}

\noindent\textbf{Popularity and User Experience Analysis.}
We investigate if app popularity correlates with the presence of immediate consent mechanisms for the user. 

\autoref{fig:pt2-boxplots} shows the install count distributions for apps with and without consent mechanisms. We bin the apps by popularity (<10K, 10K–100K, 100K–1M, 1M+ installs) and calculate consent rates within each bin.  Apps with 1M+ installs show the highest consent rate (4.98\%), while apps with <10K installs show the lowest (1.54\%). The bins corresponding to intermediate levels of popularity show consent rates of 3.87\% (10K–100K) and 3.03\% (100K–1M).
A chi-square test of independence confirms these differences are statistically significant: $\chi^2=58.17$ with three degrees of freedom, $p<0.0001$.
Furthermore, post-hoc pairwise comparisons with Bonferroni $p$-value correction show that the least popular apps differ significantly from others ($p<0.01$). We find no significant relationship between the presence of consent mechanisms and user ratings. A Mann-Whitney U test reveals $U=152K, p<0.01$, rank-biserial correlation = -0.22), indicating that users do not penalize or reward consent mechanisms through ratings.
\noindent\textbf{Key finding: Even the most popular mHealth apps exhibit alarmingly low consent rates ($<$5\%).}
This positive correlation between popularity and the presence of consent mechanisms may reflect greater regulatory scrutiny of high-profile apps, larger development teams with compliance resources, or survivor bias (i.e., non-compliant popular apps may have been removed).
However, the universally low rates also suggest systemic failures in consent implementation across the mHealth ecosystem, regardless of popularity.
%
Among the 3,346 apps without consent acquisition mechanisms, \textbf{PII collection is dominated by Google Advertising ID} (AAID, $n=$249), with minimal collection of other identifiers: email addresses (n$=$8), nearby Bluetooth MAC addresses ($n=$4), BSSID ($n=$3), and SSID ($n=$1).
No apps collected restricted identifiers (WiFi MAC, IMEI, and Device Serial Number).

\vspace{-0.2cm}
\subsection{PT3: PII Collection Beyond Consent.} 
PT3 measures whether apps with consent mechanisms collect PII beyond their disclosed scope. We compare PII mentioned in consent forms against actual collection detected through network traffic and log analysis (see \S\hspace{1pt}\ref{sec:PT3methodology}).

\begin{table}[!t]
\centering
\caption{PT3: Breakdown of PII consent and collection for apps that have a consent mechanism \textnormal{(\#Acquired: 1 for each claim an app makes to collect PII; \#Collected: 1 for each PII category found in an app's logs or network traffic; \#In gap: 1 for each PII found in an app's consent gap)}}
\renewcommand{\arraystretch}{0.75}
\begin{tabularx}{\linewidth}{>{\raggedright\arraybackslash}X c c c}
\toprule
\textbf{PII Category} & {\bf \#Acquired} & \ {\bf \#Collected } &{\bf\#In 
gap} \medskip\\
  AAID     &     31  &    21   &    17   \\
  BSSID    &      9 &      0 &     0  \\
  Device Serial Number  &      46  &     0  &     0  \\
  Email    &      98 &     1  &    1   \\
  IMEI     &      21 &     0  &    0   \\
  SSID     &     6  &     0  &     0  \\
Bluetooth Mac     &   22    &   0    &    0   \\
  \bottomrule
\end{tabularx}
\label{tab:pii-beyond-consent}
\vspace{-0.3cm}
\end{table}
Of the 253 apps displaying consent forms to the user, we find that \textbf{18 apps (7.1\%) collect PII not disclosed in their consent agreements}, with violations limited to one undisclosed PII category per app.
\autoref{tab:pii-beyond-consent} displays the mentions of a specific PII in an app's consent mechanism, and compares it against the actual collection of that PII by the app.
~
The low violation rate (7.1\%) among mHealth apps with consent mechanisms contrasts sharply with the extraordinarily high rate of complete absence of initial consent acquisition (93\%, found in our analysis of PT2).
This demonstrates that when developers implement consent, they generally align data collection with disclosure. The small sample size ($n=$ 18) limits stronger conclusions, however.

Notably, we find no evidence of apps collecting restricted identifiers (e.g., IMEI) through side channels, despite our methodology being capable of detecting such collection through network traffic and system logs, suggesting Android's security effectively restricts identifier access.


\noindent\textbf{Geographic Patterns}
For PT3, our geolocation analysis found no significant patterns, likely due to the insufficient statistical power in the sample of 18 violations. Thus, robust conclusions about regional compliance cannot be drawn.
\noindent\textbf{Popularity and User Experience Analysis} The distributions of \#installs and average rating scores indicate no clear trend relating consent gaps and popularity metrics.
This is confirmed by computing the Spearman correlation between consent gap size and these metrics: we find no statistically significant correlation (at $p=$ 0.01) between gap size, \#installs, and the average rating scores.

\vspace{-0.35cm}
\section{Discussion}

\label{sec:discussion-and-limitations}
\noindent {\em Systemic Privacy Failures in mHealth.}
Our findings reveal that mHealth apps systematically fail to protect user privacy. While similar privacy failures have been documented 
~\cite{lyons_log_2023,nguyen_share_2021,nguyen_freely_2022}, our large-scale comprehensive analysis reveals their pervasiveness in this sensitive domain. The implications may be severe: exposed PII may be linked to health status and diagnostic data, carrying the risk of discrimination and stigma.
Concerningly, even apps from well-resourced developers (e.g, ANT+, Samsung, Google) exhibit privacy violations. This suggests that the problem is not simply one of under-resourced developers failing to implement best practices, but rather systemic inattention  across the entire mHealth app ecosystem. Moreover, as the top 5\% of apps account for 92\% of total downloads, privacy issues in popular apps have a disproportionate effect.

\noindent\textit{Developer Knowledge Gap.}
The patterns we observe align with persistent findings across the literature, including developers' confusion, and lack of awareness and/or misunderstanding of privacy norms~\cite{felt2011android,xu_2019,liu_2021,nguyen_share_2021,nguyen_freely_2022}.
Our results illuminate how these knowledge gaps have disparate impacts. The issues are more prevalent for smaller developers (often relying on app store approval as stamp of correctness~\cite{alomar_developers_2022}); however larger developers still exhibit concerning lapses.
Without structural changes like clearer platform requirements and meaningful enforcement, these patterns are likely to persist regardless of developer intent.

\noindent\textit{Limitations and Scope.}
While our methodology uncovers strong evidence of privacy violations, it only captures initial app execution before initial account creation. This enables scale: unlike prior work which performs deeper analysis~\cite{neupane_data_2022,hassan_what_2023}, we analyze 3,500+ to 10,000+ mHealth apps across the three threat types. While we experimented with automated LLM-based testing including account creation steps, we found that this loop to be inefficient in terms of time and computational resources, and prone to high failure rates.

Our dynamic analysis approach introduces selection bias due to some apps failing to execute in PT2 and PT3. As we only report privacy failures observed in testing, these crashes causes us to conservatively underestimate the prevalence of privacy threats.
We focus exclusively on Google Play Store apps. While this overlooks other platforms, the Play Store dominant market position ensure meaningful coverage of the current mHealth app ecosystem.

\noindent\textit{Implications for Regulation and Research.}
Our findings suggests that voluntary compliance and app store review processes are insufficient to protect users. The current ecosystem fails users at scale, and closing this gap demands coordinated action from multiple stakeholders.
For regulators, 
requiring automated privacy auditing as a condition of market access may be desirable. For researchers, our work opens several directions, including measurements of user privacy awareness; and scalable automated analysis techniques.

\vspace{-0.2cm}

\section{Related Work not Discussed Elsewhere}
\label{sec:RelatedWork}

We discussed the most closely related publications organically through the paper. Here, we provide a brief discussion of research complementary to ours.

\noindent\textbf{App Analysis Beyond Privacy:}
Mobile app research often prioritizes risk over privacy, employing methods like risk scoring based on sets of permissions~\cite{peng_probablistic_2012,merlo_riskindroid_2017}, user expectations vis-\`{a}-vis runtime behavior~\cite{jing_riskmon_2014}, and static taint analysis (e.g., FlowDroid~\cite{arzt_flowdroid_2014}). Others focus on human factors, e.g., user security behaviors~\cite{huang_evaluating_2023} or permission comprehension~\cite{felt2012android}.
However, these work lacks the precision for privacy-centric measurement of mHealth apps, a domain that requires tailored analysis, and affects millions of users. Our approach enables automated scalable privacy assessment. 

 \noindent\textbf{Privacy Beyond Mobile Apps:} GDPR compliance frameworks dominate privacy research, spanning cloud services~\cite{mohan_analyzing_2022}, website consent management~\cite{liu_opted_2024}, web development frameworks~\cite{ferreira_rulekeeper_2023}, and identity management~\cite{sassetti_assurance_2023}. 
 These studies typically identify overt violations or dark patterns but neglect mobile ecosystems. mHealth apps pose unique challenges due to handling sensitive data, yet operating in a fragmented regulatory landscape (for challenges in assessing mHealth app privacy, see \cite{benjumea2020privacy}). Our work 
 adapts  core privacy principles 
to mobile-specific privacy threats (PT1--PT3), while avoiding reliance on user perceptions.

 \noindent\textbf{NLP for Privacy Analysis:}
NLP has emerged as a critical tool for auditing mobile app privacy practices. 
Recent work addresses regulatory alignment (e.g., mapping GDPR requirements~\cite{herwanto2024}) and risks of sensitive data leaks (e.g., detecting PII mishandling~\cite{silva2020}). Beyond compliance, researchers have identified systemic gaps between app behaviors and permission declarations (e.g.,~\cite{maji2021}).
 Subsequently, privacy-related text comprehension
was improved with the use of retrieval-augmented language models~\cite{zhang2024enhanced} and recent generative language models~\cite{duc2025,phan2024,salvi2024,zoubi2024privat5}.
Our use of NLP, however, is orthogonal to these advances, as we leverage language models to perform multi-label multiclass classification on app descriptions, where the language is more accessible to the average user, but the permission gap is harder to perceive.

\vspace{-0.3cm}

\section{Conclusion}

\label{sec:Conclusion}

Our work brings to the forefront privacy-related concerns for mHealth apps. We define three privacy threats at the point of data collection, grounded in LINDDUN framework. 
Our analysis of  these threats reveals that  69\% of apps  request permissions that are not substantiated by their stated purpose, 
only 7\% of apps present consent acquisition mechanisms to users, 
and   1 out of 14 apps that present a consent form collects data not authorized by the user. 
~
This picture is problematic for users and regulatory bodies, as privacy best practices, even when coded in laws in many jurisdictions, 
are regularly and consistently ignored by app developers. One possible avenue for future work is that of \textit{automating regulatory compliance}, i.e., building dedicated, privacy-focused tools---possibly based on what is used in ours and other measurement efforts---to automatically identify privacy threats in app behavior.

\vfill\pagebreak

\end{document}